\documentclass{kapproc}
\usepackage{t1enc}
\usepackage{procps} 
\usepackage[dvips]{graphicx}
\upperandlowercase
\kluwerbib

\begin{document}

\articletitle{The Luminosity and Metallicity Relation of distant luminous
infrared galaxies}

\author{Yanchun Liang$^{1,2}$, Fran\c{c}ois Hammer$^{2}$, Hector
Flores$^{2}$, David Elbaz$^{3}$,\\ Delphine Marcillac$^{3}$, Licai
Deng$^{1}$ and Catherine J. Cesarsky$^{4}$}

\affil{$^{1}$National Astronomical Observatories, Chinese Academy of Sciences, 
20A Datun Road,\\ $~~$Chaoyang District, Beijing 100012, P.R. China;\\
$^2$GEPI, Observatoire de Paris-Meudon, 92195 Meudon, France;\\
$^3$CEA, Saclay-Service d'Astrophysique, Orme des Merisiers, F91191,
Gif-s\^{u}r-Yvette, France;\\
$^4$ESO, Karl-Schwarzschild Strase 2, D85748 Garching bei M\"unchen, Germany
}
\email{ycliang@bao.ac.cn}

\begin{abstract}
A sample of 55 distant ($z>0.4$) luminous infrared galaxies (LIRGs)
selected from ISOCAM deep survey fields (CFRS, UDSR, UDSF) have been
studied on the basis of their high-quality optical spectra from
VLT/FORS2. We emphasize that the only way to derive O/H metal
abundances with sufficient accuracy is by deriving the extinction and
underlying Balmer absorption properly, on the basis of good-S/N
spectra of moderate resolution. Here, the extinction coefficient is
estimated via two independent methods, e.g., the Balmer-line ratio
[$A_V$(Balmer)] and the energy balance between the infrared and
H$\beta$ luminosities [$A_V$(IR)], are internally consistent, with a
median value of 2.36. Oxygen abundances [12+log(O/H)] in the
interstellar medium of the sample galaxies have been estimated from
the extinction corrected emission-line ratios, and show a range from
8.36 to 8.93, with a median value of 8.67, which is 0.5 lower than
that of the local bright disks (i.e., $L^*$) at the given magnitude. A
significant fraction of distant large disks are indeed LIRGs. Such
massive disks could have formed $\sim50$\% of their metals and stellar
masses since $z\sim1$.
\end{abstract}


\section{Introduction}

The cosmic infrared background resolved by ISOCAM shows that the
co-moving density of infrared light due to the luminous infrared
galaxies ($L_{\rm IR}\geq 10^{11} {\rm L}_\odot$; LIRGs) was more than
40 times greater at $z\sim1$ than today (Elbaz et al. 2002). The main
driver for this evolution is the emergence of luminous infrared
starburst galaxies seen by ISO at $z > 0.4$ (Flores et
al. 1999). Source counts and star-formation rates (SFRs) have been
studied for the ISO-detected objects in the $z>0.4$ Universe (Elbaz et
al. 1999, Aussel et al. 1999, Flores et al. 1999, Franceschini et
al. 2003). However, very little was known about the chemical
properties of distant LIRGs. In this paper, we present the chemical
abundances of a large sample of distant ($z>0.4$) LIRGs selected from
ISOCAM deep survey fields (CFRS, UDSR, UDSF). We estimate their oxygen
abundances in the interstellar medium (ISM) on the basis of their
high-quality optical spectra from VLT/FORS2. In particular, this is
the first time that the luminosity-metallicity (L-Z) relations of
such a large sample of distant LIRGs are obtained.

In the local Universe, metallicity correlates well with the absolute
luminosity (stellar mass) of galaxies over a wide magnitude range
(e.g., $7-9$ mag; Zaritsky et al. 1994, Richer \& McCall 1995, Telles
\& Terlevich 1997, Contini et al. 2002, Melbourne \& Salzer 2002,
Lamareille et al. 2004, Tremonti et al. 2004). In the
intermediate-redshift Universe, L-Z relations have also been obtained
for some sample galaxies. Kobulnicky \& Zaritsky (1999) found that the
L-Z relation of 14 emission-line galaxies with $0.1<z<0.5$ is
consistent with that of the local spiral and irregular galaxies. The
16 CFRS galaxies at $z\sim0.2$ studied by Liang et al. (2004a) fall
well into the region occupied by the local spiral galaxies. Kobulnicky
et al. (2003) obtained the L-Z relation of 64 intermediate-$z$
galaxies from the Deep Groth Strip Survey (DGSS). In the $0.6-0.82$
redshift bin, their galaxies are brighter by $\sim2.4$ mag, compared
to the local ($z<0.1$) field galaxies (Kennicutt 1992a,b, hereafter
K92a,b; Jansen et al. 2000a,b, hereafter J20a,b). This result was
confirmed by Maier et al. (2004). Kobulnicky \& Kewley (2004) obtained
the L-Z relation for 204 emission-line galaxies in the GOODS-N field,
and showed a decrease in the average oxygen abundance of $\sim0.14$
dex from $z=0$ to $z=1$ for the galaxies with $M_B$ from $-18.5$ to
$-21.5$. These studies contrast with the result of Lilly et
al. (2003), who found that the L-Z relation of most of their 66 CFRS
galaxies with $0.5<z<1$ is similar to that of the local galaxies from
J20b. However, Lilly et al. (2003) assumed a constant $A_V=1$ in order
to account for dust extinction. In this study, we investigate the L-Z
relation for LIRGs at $z>0.4$ deteced by ISO, after taking into
account their underlying stellar absorption and dust extinction
properties in detail.

Throughout this paper, a cosmological model with $H_0=70$ km s$^{-1}$
Mpc$^{-1}$, $\Omega_{\rm M}=0.3$ and $\Omega_\Lambda=0.7$ has been
adopted. $M_B$ is given in the AB magnitude system.

\section{Sample selection, observations and data reduction}

The sample galaxies were selected from three ISO deep survey fields:
the CFRS 3h field, the Ultra Deep Survey Rosat (UDSR) and the
Ultra Deep Survey FIRBACK (UDSF) fields. In the CFRS 3h field, 70
sources were detected, with their 15 $\mu$m fluxes in the range of
$170-2100 \,\mu$Jy (Flores et al. 2004, 2005, in prep.). The UDSR field
refers to the Marano field, which is a deep ROSAT field. FIRBACK is a
deep survey conducted with the ISOPHOT instrument aboard ISO, at an
effective wavelength of 175 $\mu$m. For the UDSR and UDSF fields, very
deep ISOCAM follow-up observations have been obtained (Elbaz et
al. 2005, in prep.) reaching flux limits three times lower than for
the CFRS field.

In total, 105 objects were selected for our VLT/FORS2 spectral
observations with R600 and I600, at a resolution of 5 {\AA} and
covering $5000 - 9200$ {\AA}. The slit width was 1.2 arcsec. Spectra
were extracted and wavelength calibrated using the {\sc iraf}
package. Flux calibration was done using 15-minute exposures of 3
photometric standard stars per field.

The redshift distribution of the sample galaxies shows a median value
of 0.587, which is consistent with the results in some other ISOCAM
survey fields (Flores et al. 1999, Aussel et al. 1999, Franceschini et
al. 2003). Fifty five of the redshift-identified objects are distant
LIRGs with $z>0.4$.

The IR luminosities (and deduced SFRs) were calculated using the
procedure given in Elbaz et al. (2002). The inferred IR luminosity
($8-1000 \,\mu$m) of the ISOCAM $15\mu$m-detected objects with $z>0.4$
shows a similar distribution as the local IRAS sample from Veilleux et
al. (1995) and Kim et al. (1995), with a median value of $\log(L_{\rm
IR}/{\rm L}_\odot)=11.32$.

\section{A robust estimate of the extinction coefficient}

The extinction inside a galaxy can be derived using its Balmer
decrement. For the $z>0.4$ galaxies, H$\alpha$ is shifted to the near
infrared, so that H$\gamma$/H$\beta$ can be used to estimate the dust
extinction. The stellar absorption under the Balmer lines was
estimated from synthesized stellar spectra obtained using the stellar
spectral library of Jacoby et al. (1984). We adopt the interstellar
extinction law of Fitzpatrick (1999) with $R=3.1$, and Case B
recombination, with a density of 100\,cm$^{-3}$ and a temperature of
10,000\,K to estimate the dust extinction. Subsequently, the
extinction-corrected Balmer emission lines (either H$\beta$ or
H$\alpha$) were used to estimate the SFRs by adopting the calibrations
from Kennicutt (1998) based on a Salpeter (1955) initial mass function
(IMF; 0.1 and 100 M$_\odot$ mass cut-offs). The derived SFR$_{\rm
Balmer}$ could be compared with the SFR$_{\rm IR}$ obtained from the
infrared flux. The two SFRs are consistent, and their median values
are 28 and 31 M$_\odot$ yr$^{-1}$, respectively. This kind of
consistency was also confirmed by Hopkins et al. (2003) for the SDSS
galaxies, and by Kewley et al. (2002) for the Nearby Field Galaxies
Survey sample of J20b.

Because of the large uncertainties related to the measurements of the
H$\gamma$ line, we need to verify the quality of our derived
extinction values. This can be done assuming that the infrared data
provide a robust SFR estimate for IR-luminous galaxies (Elbaz et
al. 2002, Flores et al. 2004). We estimate a new dust extinction
coefficient, $A_V$(IR), by comparing SFR$_{\rm IR}$ and SFR$_{2.87{\rm
H}\beta}$, the energy balance between the IR and H$\beta$
luminosities. Figure\,1 shows that the derived $A_V$(IR) is
consistent with $A_V$(Balmer) for most galaxies, most of them falling
in the $\pm 0.64$ rms discrepancy. The derived median value of
$A_V$(IR) is 2.36 for the $z>0.4$ galaxies.

\begin{figure}[ht] 
  {\centering
   \includegraphics[bb=26 152 572 702,width=4.8cm,clip]{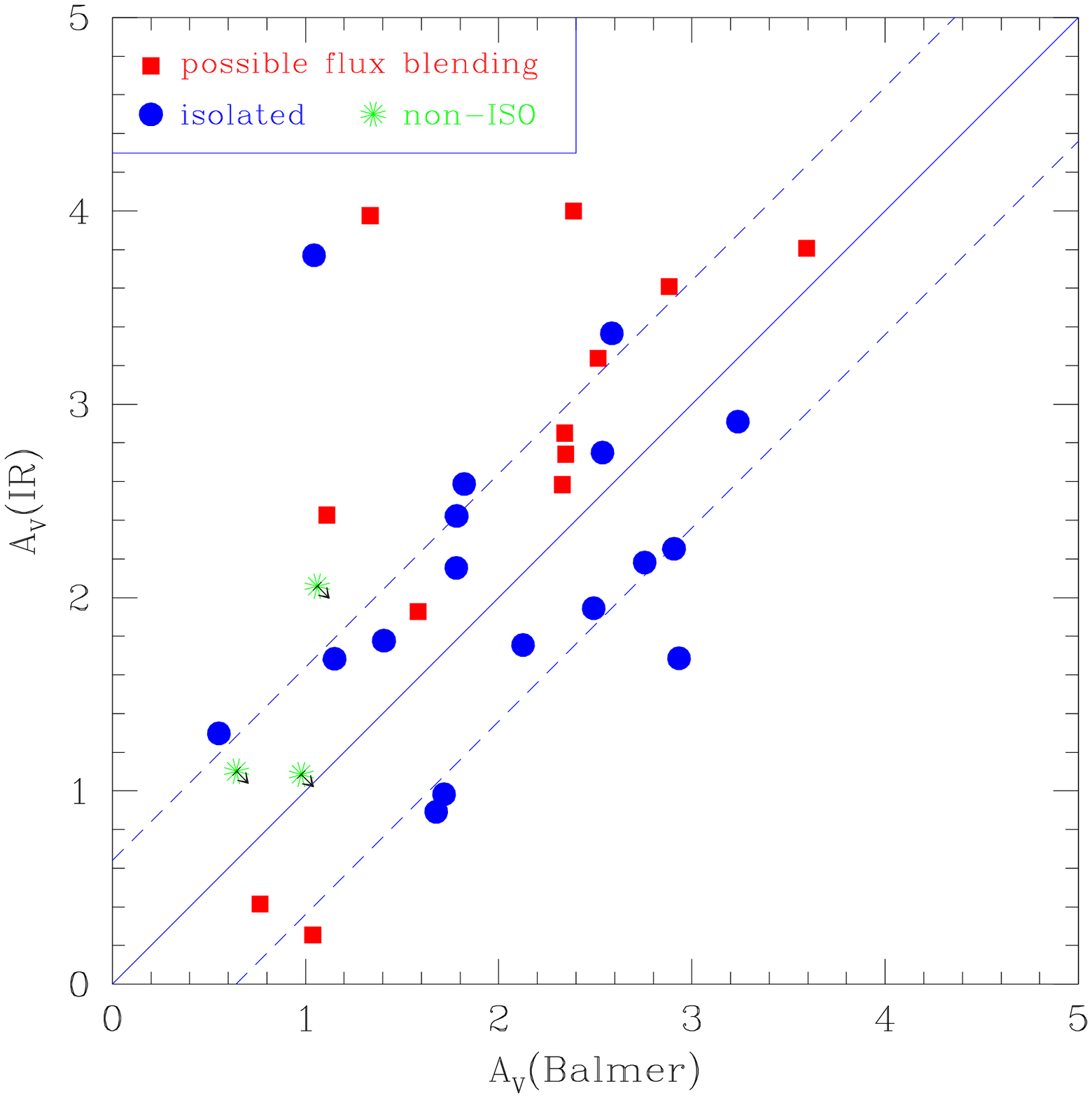} 
\caption{The relation between the extinction values derived from the
Balmer decrement [$A_V$(Balmer)], and from the energy balance between
the IR radiation and the optical H$\beta$ emission-line luminosities
[$A_V$(IR)]. They are consistent with each other. The two dashed lines
refer to the results with $\pm 0.64$ rms.} }
\end{figure}

\section{Abundances in the interstellar medium and the 
luminosity-metallicity relation}

The chemical properties of the gas and stars in a galaxy are like a
fossil record chronicling its history of star formation and its
present evolutionary status. The high-quality optical spectra from
VLT/FORS2 make it possible to obtain the chemical abundances of the
ISM for these distant LIRGs. This will be the first data set of
chemical abundances of a large sample of distant LIRGs.

The oxygen abundances of the distant LIRGs were estimated via the
``strong-line'' method ($R_{23}$ and $O_{32}$), using the calibration
of Kolbinicky et al. (1999) for the metal-rich branch. The derived
12+log(O/H) values of the galaxies range from 8.36 to 8.93, with a
median value of 8.67.

Figure\,2a compares the L-Z relation for the LIRGs to that of local
disks (K92b, J20b), which are restricted to moderately star-forming
[EW(H$\beta$) $<20$ {\AA}] galaxies, following Kobulnicky et
al. (2003). Here we did not include other local samples selected based
on their UV or H$\alpha$ emission; they mostly include low-luminosity
(low-mass?) systems in the local Universe (Contini et al. 2002,
Lamareille et al. 2004, Melbourne \& Salzer 2002). We could not
compare directly with the large sample of SDSS galaxies given by
Tremonti et al. (2004) because of the large scatter of the data.

The distant LIRGs exhibit $\sim0.3$ dex ($\sim50$\%) lower oxygen
abundances than local star-forming galaxies at the given magnitude
(the median value). \mbox{{\sc P}\hspace{-0.04cm}~\'{\sc
\hspace{-0.19cm}egase2}} models (Fioc \& Rocca-Volmerange 1999)
predict a total mass ranging from $10^{11}$ M$_\odot$ to $\le 10^{12}$
M$_\odot$ for the LIRGs, which can be twice the stellar masses of
distant LIRGs ($1.4 \times 10^{10} - 2.9\times10^{11}$ M$_\odot$)
derived by Zheng et al. (2004) on the basis of $K$-band
luminosities. About 36\% of LIRGs are large disks; a similar fraction
(about 32\%) is estimated from the sample of large disks of Lilly et
al. (1998). These massive LIRGs have high SFRs. The time-scale to
double their stellar masses $T_{\rm SF}$ can be short, i.e., $0.1-1$
Gyr. Such massive disks could have formed $\sim50$\% of their metals
and stellar masses since $z\sim1$. Hammer et al. (2004, 2005) have
investigated whether the LIRG properties could be related to recent
and significant star formation in massive galaxies, including spirals.

These distant LIRGs were also compared with two samples of galaxies in
a similar redshift range, taken from Kobulnicky et al. (2003)
[EW(H$\beta$) \mbox{$<20$ {\AA}]} and Lilly et al. (2003). Kobulnicky et
al. (2003) estimated the O/H values using the $R_{23}$ and $O_{32}$
parameters obtained from the corresponding equivalent widths of the
emission lines, which are believed to be less affected by dust
extinction (Kobulnicky \& Phillips 2003). The metallicities of their
galaxies are similar to ours, but the galaxies are fainter at a given
metallicity. This discrepancy in $M_B$ reflects that our sample
galaxies are brighter and possibly more massive than the rest-frame
blue-selected sample of DGSS galaxies (see Liang et al. 2004b for
details).

The comparison with Lilly's sample helps us to understand the strong
effects of dust extinction on the derived oxygen abundances of
galaxies. In Fig.\,2b, Lilly's sample has been restricted the galaxies
in the CFRS 3h and 14h fields, the two fields surveyed by ISOCAM. The
10 ISO galaxies (the solid triangles) among the 42 sample galaxies
show a median value of 12+log(O/H) = 8.98, which is $\sim0.3$ dex
higher than the median value of our distant LIRGs. Indeed, Lilly et
al. (2003) assumed a constant extinction of $A_V=1$ for all their
galaxies, which can underestimate the average extinction for LIRGs
(which has a median value of 2.36), and consequently leads to
underestimated [O{\sc ii}]$\lambda3727$/H$\beta$ ratios, and hence to
overestimated oxygen abundances. To estimate the chemical abundances
of such distant LIRGs, underlying stellar absorption of Balmer lines
and dust extinction should be considered carefully.

\begin{figure*} 
{ \centering
  \includegraphics[bb=72 435 408 571,width=10.6cm,clip]{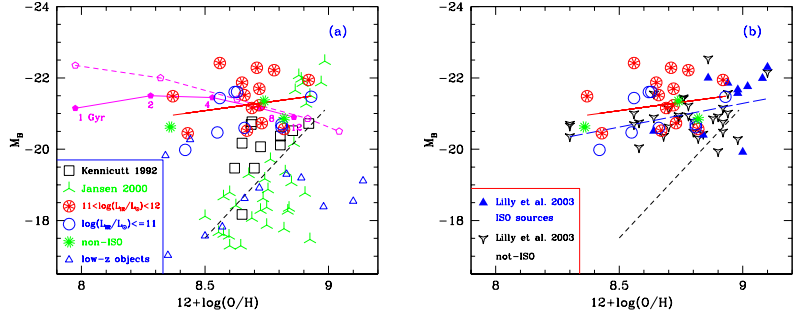} 
\caption{The $M_B$-metallicity relation of our distant LIRGs (with a
 typical uncertainty of 0.08 dex on metallicity), compared with other
 samples: {\em (a)} with the local galaxies from K92b and J20b;
 \mbox{{\sc P}\hspace{-0.04cm}~\'{\sc \hspace{-0.155cm}egase2}} infall
 models are superimposed, assuming a total mass of $10^{11}$ M$_\odot$
 and infall times of 5 Gyr and 1 Gyr (solid and dashed lines with
 pentagons, respectively); {\em (b)} with the galaxies of Lilly et
 al. (2003). The linear least-squares fits to the samples are also
 given.}}
\end{figure*}

\begin{acknowledgments}
Yanchun Liang wishes to thank the Natural Science Foundation of China
(NSFC) for travel support to attend this conference, also thank the 
NSFC support under No.10403006. We thank Dr. Richard de Grijs for his 
help to improve the English description.
\end{acknowledgments}

\end{document}